\documentclass{article}
\usepackage{ijcai20}

\usepackage{times}
\usepackage{soul}
\usepackage{url}
\usepackage[hidelinks]{hyperref}
\usepackage[utf8]{inputenc}
\usepackage[small]{caption}
\usepackage{graphicx}
\usepackage{amsmath}
\usepackage{amsthm}
\usepackage{booktabs}
\usepackage{algorithm}
\usepackage{algorithmic}
\usepackage{csquotes}

\usepackage{url}

\usepackage{breakurl}

\title{Narratives and Needs: Analyzing Experiences of Cyclone Amphan Using Twitter Discourse}

\author{
Ancil Crayton$^1$
\and
João Fonseca$^3$\and
Kanav Mehra$^5$\and
Michelle Ng$^2$\and
Jared Ross$^1$\and
Marcelo Sandoval-Castañeda$^4$\And
Rachel von Gnechten$^2$
\affiliations
$^1$Booz Allen Hamilton, 
$^2$International Water Management Institute, 
$^3$NOVA Information Management School (NOVA IMS),
$^4$New York University Abu Dhabi,
$^5$Independent Researcher
\emails
ancil.crayton@ucdconnect.ie, jpfonseca@novaims.unl.pt,
\{jaredrossj, kanav.mehra6\}@gmail.com,
\{m.ng, r.vongnechten\}@cgiar.org,
marcelo.sc@nyu.edu
}

\begin{document}

\maketitle

\begin{abstract}
  People often turn to social media to comment upon and share information about major global events. Accordingly, social media is receiving increasing attention as a rich data source for understanding people's social, political and economic experiences of extreme weather events. In this paper, we contribute two novel methodologies that leverage Twitter discourse to characterize narratives and identify unmet needs in response to Cyclone Amphan, which affected 18 million people in May 2020.
\end{abstract}

\section{Introduction}
With wind speeds gusting up to 200 kilometres per hour, Cyclone Amphan was the first super cyclone to form in the Bay of Bengal since 1999 \cite{introNasa}. It made landfall in West Bengal, India on May 20, 2020 before tracing a destructive path northward to Bangladesh \cite{introForbes}. Along the way, Cyclone Amphan damaged nearly 3 million houses, 18,000 square kilometres of agricultural lands and 449,000 electric poles, leaving 18 million affected people in its wake \cite{introRedcross,introState}. Cyclone Amphan was also the costliest cyclone in the history of the North Indian Ocean. India reports 13.2 billion USD in damages in the state of West Bengal alone \cite{introState}.

Extreme weather events are expected to increase in magnitude and frequency due to the impacts of climate change; and Cyclone Amphan is one such example. The Bay of Bengal’s unprecedentedly high sea surface temperature, which is linked to anthropogenic climate change, likely contributed to Cyclone Amphan’s speed and energy \cite{introIwmi}. Unfortunately, warming ocean temperatures will intensify more cyclones and hurricanes in the future — both in the Bay of Bengal and beyond. Thus, it is imperative to develop and refine approaches for responding to extreme weather events that draw upon all available tools.

In the case of Cyclone Amphan, response efforts were complicated by COVID-19. For instance, in addition to the typical heavy rains and obstructed roads, responders had to cope with restricted mobility due to India’s nationwide lockdown, limitations on shelter capacities due to social distancing measures and the need to obtain, use and distribute personal protective equipment \cite{introState,introRedcross}. On-the-ground response efforts by governments, disaster relief organizations and civil society are no doubt crucial and life-saving following extreme weather events. Could online data serve as an additional tool to supplement on-the-ground efforts, particularly when they are hindered? Although online data cannot paint a complete or representative picture of offline realities, it could help fill knowledge gaps when there are challenges reaching affected people, such as those caused by COVID-19.

We took Cyclone Amphan as our use case in exploring the potential for Twitter content to target relief efforts in response to extreme weather events. We first aimed to characterize how collective knowledge about Cyclone Amphan was produced on Twitter. Twitter is a decentralized microblogging platform, meaning that anyone from anywhere in the world can add their commentary to an issue — thus influencing its narrative and adding layers of interpretation to on-the-ground realities. After exploring who and what is shaping the narratives around Cyclone Amphan, we aimed to answer: Can Twitter content help identify unmet needs of people affected by Cyclone Amphan? If so, how?

\section{Natural Disasters and Social Media}
Social media platforms serve as massive repositories of real-time situational and actionable data during man-made or natural emergencies, such as extreme weather events. For instance, people can use social media to organize volunteer or donation campaigns in support of on-the-ground relief efforts or directly contact relevant organizations via their official social media accounts \cite{Imran2015}. Thus social media posts regarding extreme weather events vary broadly from people sharing personal experiences and opinions to emergency response agencies posting updates, warnings and information about relief efforts.

Furthermore, social media content can be used to characterize people's experiences of extreme weather events and trace how narratives took shape through collective knowledge production. Several existing studies focus on developing efficient and scalable methods for extracting important, actionable information from social media content using a range of techniques based on natural language processing (NLP), text mining and network analysis \cite{Imran2013,Imran2013a,Imran2014}. A large volume of work in this domain focuses on developing classifiers to categorize tweets as informative or uninformative using learning-based approaches \cite{Alam2018,Hernandez-Suarez2019,Zhang2016} or matching-based approaches \cite{To2017,Mehra2017}. Due to the unavailability of large labeled datasets, unsupervised or semi-supervised learning methods \cite{Alam2018,Zhang2016,Arachie2020} are preferred over supervised learning based approaches, which tend to not generalize well across different crisis events. To the best of our knowledge, we provide the first usage of zero-shot text classification to assign tweets to multiple classes relevant to the impacts of extreme weather events. This method not only utilizes the abundantly available unlabeled Twitter data (tweets, or microblogs) and circumvents the need for supervised learning, but also facilitates the application of a pre-trained language model that makes the overall method generalizable across different crisis events. 

Another active direction of research is the identification of important sub-events within a large-scale extreme weather event for efficient management of relief efforts \cite{Rudra2018}. Recent work proposes an unsupervised learning-based framework using semantic embeddings of noun-verb pairs from tweets to detect sub-events \cite{Arachie2020}. Both extractive and abstractive methods of text summarization have been studied to effectively summarize the huge volumes of microblogs posted during emergency events \cite{Rudra2015,Rudra2016,Mehra2017,Dutta2018}. Researchers have also explored methods to classify microblogs posted during natural disasters by sentiment and information utility \cite{Ragini2018,Zhang2016}. However, our paper aims to develop a more holistic analysis of social media content for disaster management. We combine methods of sentiment analysis and classify tweets into multiple relevant classes based on the information they convey. Besides adding more context to the tweets as well as the classes they are placed in, our methodology explores the emotional spectrum associated with the classes. 

While the content analysis of microblogs during emergency events has been extensively discussed, the analysis of social network structure along with patterns of user behaviour on Twitter during extreme weather events is still relatively unexplored \cite{Pourebrahim2019}. In this paper, we intend to address this gap by studying the embedded social network structure and comparing user behavior across different sets of users to clearly distinguish between individuals who shaped the dominant narrative and those who were marginalized.

\section{Approach}
\subsection{Dataset}
We extracted around 470,000 tweets using the Twitter API. We targeted tweets from May 1st, 2020 to June 15th, 2020, which cover the build-up through the aftermath of Cyclone Amphan. The main languages targeted in our query were English, Odia, Hindi and Bengali, but some tweets in other languages were extracted as well. We also filtered out terms related to other catastrophes happening in the area during the same time period, such as Cyclone Nisarga, which hit the Indian subcontinent at the beginning of June.

\subsection{Preprocessing data}
Given the multi-language approach of our query, the first step in the preprocessing pipeline is to translate non-English tweets into English using the Google Translate API. The Twitter API identifies and tags the language of each Tweet. We take advantage of this attribute such that only non-English tweets are translated, thus reducing computational costs.

The next step is to remove URLs and reserved words from the content of the tweets. This includes hashtags, emojis and words like \enquote*{RT} or \enquote*{FAV}. Then we remove all remaining punctuation and change any uppercase letters to lowercase. As a final step, we remove all stop words found in the text of the tweets, following NLTK's stop words list for English \cite{Loper02nltk:the}, and lemmatize the remaining words using NLTK's WordNetLemmatizer.

\subsection{Feature Extraction}
The next goal is to extract information from the Twitter data. This process is divided into 4 independent tasks: 

\subsubsection{Sentiment analysis}
The extraction of data information regarding the writer's sentiment is a common, well-studied NLP task. In this work, the sentiment of each tweet is not known beforehand. Therefore, we leverage the Valence Aware Dictionary and Sentiment Reasoner (VADER) model \cite{Hutto2014} to capture the sentiment of each tweet. VADER determines the sentiment of a document through a rule-based approach using a sentiment lexicon (i.e., a list of lexical features labelled according to their semantic orientation) to determine how positive/negative a specific tweet is \cite{Hutto2014}. It is advantageous in the context of this paper as it was developed for social media text.

\subsubsection{Point-of-view extraction}
We use point-of-view extraction to classify tweets as first person, second person or third person, with the goal of determining whether the user experienced Cyclone Amphan personally. It is done by iterating over all tokens in the tweet's text and identifying any word matching a list of pronouns mapped to first, second or third person speech. Examples of first person speech could be tokens such as `I`, `my' and `our'; for second person, this may include `you' or `your'; and, finally, for third person, it may include `them', `they' or `it'.

Our classification strategy is to assign a tweet to be first, second or third person based on the order of precedence, respectively. Therefore, if the tweet contains any first person pronouns, it is designated to be first person point of view. If the tweet contains second person pronouns, but does not contain any first person pronouns, it is assigned to be second person. Finally, if the tweet contains third person pronouns, but does not contain second or first person pronouns, it is classified as third person.

\subsubsection{Zero-shot text classification}

\begin{figure}
  \includegraphics[width=\linewidth, trim=0 50 0 75]{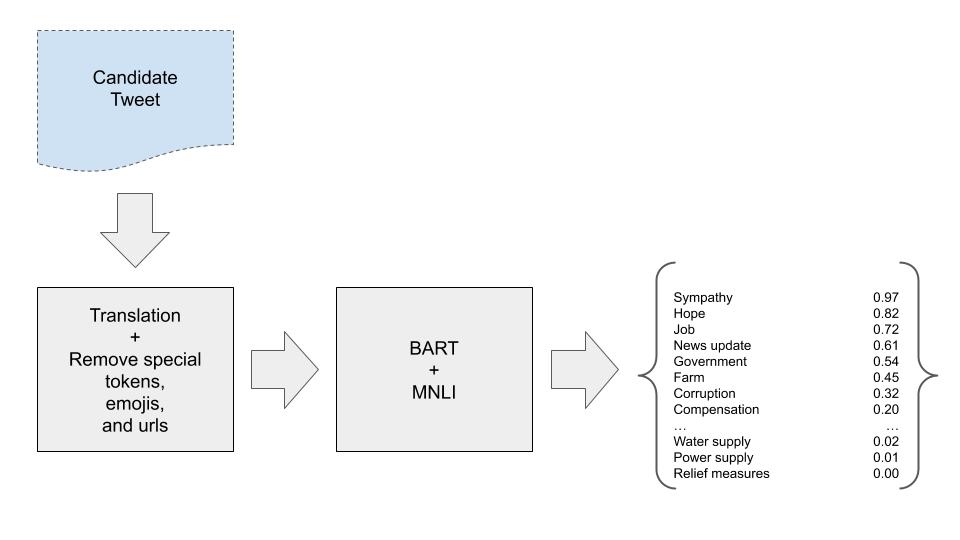}
  \caption{Zero-shot text classification pipeline.}
  \label{fig:zstcpipeline}
\end{figure}

In order to extract critical information and derive actionable insights from large volumes of social media content generated during extreme weather events, it is imperative to effectively categorise the microblogs into distinct classes.

Due to the unavailability of large annotated training sets, we present a novel application of zero-shot text classification for a multi-label classification of tweets that is not only scalable but also generalizable across different crisis events. Researchers have proposed an approach of using a pre-trained NLI (Natural Language Inference) sequence-pair classifier as a zero-shot text classifier \cite{yin2019benchmarking}. The model considers a sequence input (tweet) as the premise and each candidate topic label as a hypothesis. For our purpose, we use the zero-shot classification pipeline implementation available in the Transformers package that uses a large BART model \cite{lewis2019bart} pre-trained on the MNLI dataset \cite{N18-1101}. This pipeline is shown in Figure \ref{fig:zstcpipeline}.
    
After experimenting with different combinations, we settle on a comprehensive set of $N$ specific labels that cover a wide variety of information: \{`sympathy', `criticism', `hope', `job', `relief measures', `compensation', `evacuation', `ecosystem', `government', `corruption', `news updates', `volunteers', `donation', `cellular network', `housing', `farm', `utilities', `water supply', `power supply', `food supply', `medical assistance', `coronavirus', `petition', `poverty', `assistance required’\}.
    
The entire set of unique tweets is fed to the zero-shot classifier after basic preprocessing steps outlined in section 3.2. However, to preserve more text and conform the data closer to the training data of the BART model, we omit the case folding, stopword removal and lemmatization steps for this task. The classifier yields a confidence score ranging between 0 and 1 for every tweet-label pair. To ensure minimum overlap and maintain exclusivity, each tweet is assigned to a topic label if the confidence score associated with the pair is above a certain threshold, say $\alpha$. We experiment with a set of values for $\alpha$ and observe the best results with $\alpha = 0.7$.

\subsubsection{User and Tweet embeddings}~\label{sec:embeddings}
Embeddings are vector representations of either words, documents (tweets) or a set of documents (the user). They allow the conversion of non-numerical data (text) into an $n$-dimensional space, where the relationships among words, tweets and/or users is preserved. There are many methods directed toward vector representation of words. Amongst the most popular methods is one-hot encoded representations, distributed representations, Singular Value Decomposition, continuous bag of words and skip-gram model.

Tweet vectorization is done using a skip-gram model otherwise known as the Doc2Vec algorithm \cite{Le2014}. This choice was motivated by not only its popularity and computational efficiency, but also its capacity to maintain a logical spatial structure among tweets, both regarding the tweets' corpus and their underlying topic and sentiment. The Doc2Vec model is trained using unique tweets and replies in order to avoid the bias toward highly retweeted tweets that would come from keeping duplicate text. This results in a model trained on approximately 113,000 documents over 50 epochs. Tweets containing rare words in the dataset's corpus (i.e., words appearing twice or less) are rejected for training. The output are 200-dimensional embeddings for each tweet in our dataset.

The user embeddings are based on the tweet embeddings, with a process that involves averaging the tweets/retweets belonging to a given user \cite{Hallac2019}. This allows the analysis of the type of discourse and opinion shared among users in a 200-dimensional space.

\subsection{Analysis}

The extracted information is then combined to address our research questions as follows.

\subsubsection{Identifying narratives and influential users}

\begin{figure}
  \includegraphics[width=\linewidth, trim=0 50 0 75]{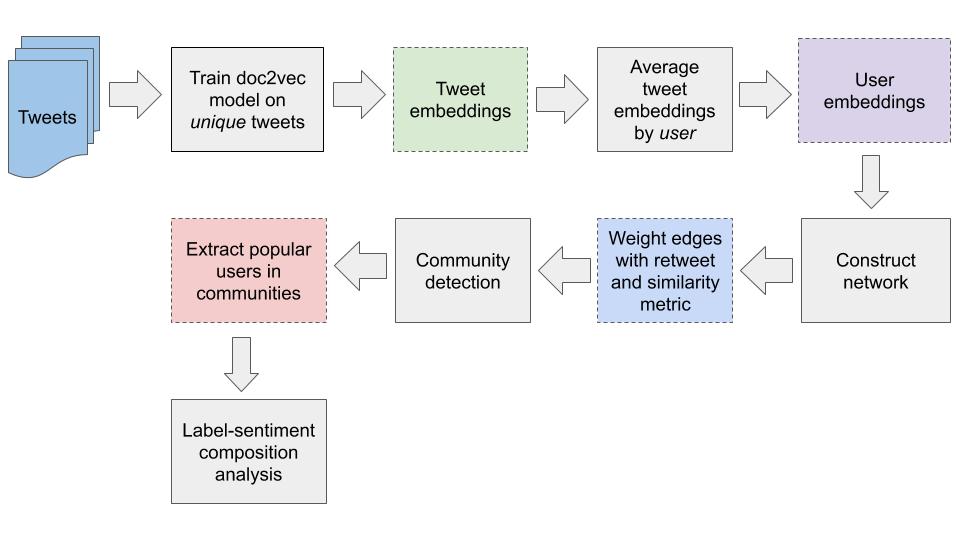}
  \caption{Pipeline for identifying the narrative and who's shaping it.}
  \label{fig:rq2pipeline}
\end{figure}

\begin{figure}
  \includegraphics[width=\linewidth, trim=4 4 4 4,clip]{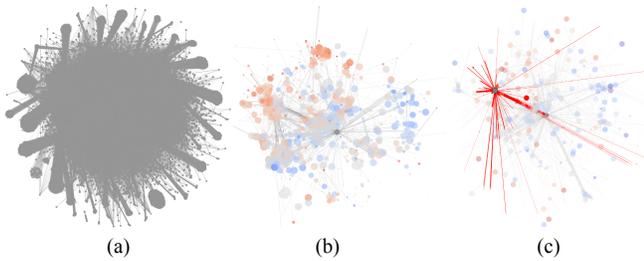}
  \caption{Twitter's user network, based on tweets related to Cyclone Amphan. Legend: (a) Complete user network, (b) Sampled user network, node size varies according to the number of tweets related to Amphan, (c) Sampled user network, node size varies according to the number of followers, with influential user Narendra Modi (India's Prime Minister) highlighted. Both figures (b) and (c) are sampled to 1000 edges.}
  \label{fig:network_results}
\end{figure}

The pipeline for identifying narratives and influential users in the dataset is shown in Figure \ref{fig:rq2pipeline}. We address this question through the usage of user vectors as described in subsection \ref{sec:embeddings} as a means of positioning users in a two dimensional space. The projection of the 200-dimensional embeddings was done using t-SNE \cite{VanDerMaaten2008}, resulting into 2-dimensional coordinates used to position each user (i.e., nodes) in the network graph. The network's edges are assigned based on the number of retweets and/or replies among users, which are then weighted by dividing this number with the users' euclidean distance (using the original 200-dimensional embeddings). These users can now be grouped into different communities using two different methods: 1) discourse-based, where the clustering is done on the embedding features and 2) community-based, done through network clustering methods. For both clustering methods, the most popular users within each cluster are identified based on centrality measures and the number of followers the user has. Lastly, we analyze these users' discourse based on labels and average sentiment associated to these users. Figure \ref{fig:network_results}a depicts the full user network, whereas Figures \ref{fig:network_results}b and \ref{fig:network_results}c are random samples of 1000 edges.

An interesting initial finding of our analysis is that the user embeddings seem to be consistent with the sentiment analysis. Users with a positive mean sentiment score tend to be on the top-left region of the graph, whereas users with a negative mean sentiment score tend to be on the opposite region. 

\subsubsection{Identifying negative experiences and unmet needs}

\begin{figure}
  \includegraphics[width=\linewidth, trim=0 50 0 50]{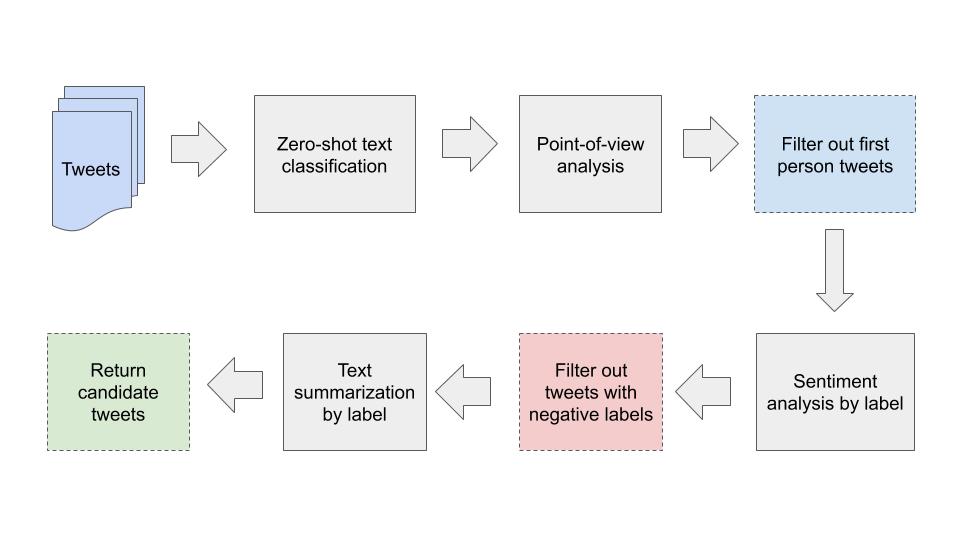}
  \caption{Pipeline for identifying negative experiences and unmet needs.}
  \label{fig:rq1pipeline}
\end{figure}

The pipeline for identifying negative experiences and unmet needs is outlined in Figure \ref{fig:rq1pipeline}. The first step is to identify the topics discussed in each tweet by assigning labels via zero-shot text classification. In order to only analyze tweets from users who were personally affected by Cyclone Amphan, the data is then filtered to include only first person tweets using the point-of-view analysis. At this point, we determine to which topic each affected individual is most sensitive. The sentiment analysis results enables understanding of whether the individual's view is either positive or negative.

We narrow down our focus to the labels that have a median negative sentiment with the assumption that negative experiences are more likely to suggest unmet needs. We then report representative tweets using extractive summarization techniques to identify the dominant themes within each label.

Our summarization method expands upon recent work that proposes an unsupervised graph-based summarization algorithm specifically for microblogs using a tweet-similarity graph over the tweet vectors generated from Doc2Vec embeddings \cite{Dutta2018}. We present a slight modification by choosing representative tweets returned from the resulting connected components based on a (maximum) score, which takes into account centrality and tweet significance as follows:
\begin{equation}
Score = C + S,
\end{equation}

where $C$ represents the degree centrality and $S$ is the node size that is calculated as $S = log(tweet\_length)$ where $tweet\_length$ is the number of tokens in the tweet.

We create $K$-length summaries for labels with negative median sentiment scores, where $K$ represents the number of representative tweets resulting from the text summarization method. The exact number to be considered can be adjusted by the researcher, policymaker or relief organization interested in learning about the experiences.

A sample of our initial results are reported in Table \ref{table:unmetneedsexample}. These three tweets are selected through the text summarization algorithm on first-person tweets in the `housing' label, which has a negative median sentiment score. Our approach allows us to identify cases of home damage, including flooding and major destruction specifically in the Siddha Galaxia Oceania and Khejuri II blocks of Kolkata, India, respectively.

%\begin{figure}
%  \includegraphics[width=\linewidth, trim=0 50 0 75]{figures_unmetneeds_example.jpg}
%  \caption{Sample of results from the negative experience and unmet needs analysis. These are 3 sample tweets that result from the text summarization over first-person tweets in the `housing' label, which has a negative median sentiment score. In this model, we set $K=50$.}
%  \label{fig:unmetneedsexample}
%\end{figure}

\begin{table}
\scalebox{0.75}{\tiny \begin{tabular}{|p{5.25cm}|p{5.25cm}|} 
\hline \multicolumn{1}{|c|}{\textbf{Full Text}} & \multicolumn{1}{|c|}{\textbf{Location}}\\
\hline
  @siddhagroup Plz don't fool people. We r residents of Siddha Galaxia Oceania block. We r suffering from poor quality windows, bedrooms of residents flooded during Amphan cyclone. Lifts are not working since Amphan cyclone. No update from Siddha when the lifts will be repaired. Shame on u. &
  Kolkata, India \\ \hline
@HYDTP @TelanganaDGP I have applied for epass to travel from Hyderabad to Kolkata on 23rd May but the status is still showing pending...it is very urgent as my house is damaged by amphan cyclone...can it be approved early or can the process be fast tracked? &
  Hyderabad, India \\ \hline
@PMOIndia @narendramodi Khejuri Block II in East Medinipur District,West Bengal is completely deatroyed caused to Amphan Cyclone Yesterday. Almost 250 Homes has been destroyed completely. I request to all Administrator to look into this area so that Khejuri Block II gets Proper help this time atleast. https://t.co/rcDUnf552W &
  Kolkata, India \\ \hline
\end{tabular}}
\caption{Sample of tweets from the negative experiences and unmet needs analysis. These result from text summarization over first-person tweets in the `housing' label, which has a negative median sentiment score. We set $K=50$.}
\label{table:unmetneedsexample}
\end{table}

\section{Conclusion}
In this paper, we contribute two methodologies for analyzing Twitter content to characterize experiences of Cyclone Amphan, including the identification of unmet needs and the collective production of narratives. We recognize that social media does not provide a complete or representative picture of extreme weather events, especially in low-resource environments where people may not have access to technology. For instance, in 2019, Internet penetration in West Bengal, India was at 29\%; of rural Internet users in India, 72\% were male \cite{conclIAMAI}. While these methodologies should consequently not be used alone, they can supplement existing on-the-ground efforts, particularly when in-person needs assessments are hindered, e.g. due to COVID-19. We anticipate these methodologies to be helpful for policy-makers, disaster relief organizations, researchers and members of civil society who wish to leverage an additional tool to better their understanding of the impacts of extreme weather events and how to focus their efforts.

% WANT THIS ALL ON NEW PAGE
\section*{Acknowledgements}
This project was completed as part of the Data Science for Social Good (DSSG) Solve for Good program. We would like to thank Andrew Bell and Jessica Toth, members of the DSSG Solve for Good team, for their assistance during this project. We would also like to thank Simon Langan at the International Water Management Institute for his support of and feedback on our work. Finally, we thank the International Water Management Institute for graciously funding data collection from the Twitter API.

\section*{Disclaimer}
The views expressed in this paper are those of the authors and not of their affiliations.

\bibliographystyle{named}
\bibliography{main} % SWAP OUT WITH OUR BIBLIOGRAPHY

\end{document}